\documentclass[12pt, draftcls, onecolumn]{IEEEtran}
\usepackage{setspace}
\usepackage{amsmath,amsfonts}
\usepackage{algorithmic}
\usepackage{algorithm}
\usepackage{array}
\usepackage[caption=false,font=normalsize,labelfont=footnotesize,textfont=sf]{subfig}
\usepackage{textcomp}
\usepackage{stfloats}
\usepackage{url}
\usepackage{color}
\usepackage{verbatim}
\usepackage{graphicx}
\usepackage{cite}
\begin{document}
\title{\Large A Tutorial on Holographic MIMO Communications—Part I: Channel Modeling and Channel Estimation}
\author{Jiancheng An,~\IEEEmembership{Member,~IEEE}, Chau Yuen,~\IEEEmembership{Fellow,~IEEE},\\Chongwen Huang,~\IEEEmembership{Member,~IEEE}, M\'erouane Debbah,~\IEEEmembership{Fellow,~IEEE},\\H. Vincent Poor,~\IEEEmembership{Life Fellow,~IEEE}, and Lajos Hanzo,~\IEEEmembership{Life Fellow,~IEEE}\\
\emph{(Invited Paper)}
\thanks{This research is supported by the Ministry of Education, Singapore, under its MOE Tier 2 (Award number MOE-T2EP50220-0019). This work was supported by the Science and Engineering Research Council of A*STAR (Agency for Science, Technology and Research) Singapore, under Grant No. M22L1b0110. The work of Prof. Huang was supported by the China National Key R\&D Program under Grant 2021YFA1000500, National Natural Science Foundation of China under Grant 62101492, Zhejiang Provincial Natural Science Foundation of China under Grant LR22F010002, Zhejiang University Global Partnership Fund, Zhejiang University Education Foundation Qizhen Scholar Foundation, and Fundamental Research Funds for the Central Universities under Grant 2021FZZX001-21. H. V. Poor would like to acknowledge the financial support of the U.S. National Science Foundation under Grant CNS-2128448. L. Hanzo would like to acknowledge the financial support of the Engineering and Physical Sciences Research Council projects EP/W016605/1 and EP/X01228X/1 as well as of the European Research Council's Advanced Fellow Grant QuantCom (Grant No. 789028). \emph{(Corresponding author: Chau Yuen.)}}
\thanks{J. An is with the Engineering Product Development Pillar, Singapore University of Technology and Design, Singapore 487372 (e-mail: jiancheng\_an@sutd.edu.sg). C. Yuen is with the School of Electrical and Electronics Engineering, Nanyang Technological University, Singapore 639798 (e-mail: chau.yuen@ntu.edu.sg). C. Huang is with College of Information Science and Electronic Engineering, Zhejiang-Singapore Innovation and AI Joint Research Lab and Zhejiang Provincial Key Laboratory of Info. Proc., Commun. \& Netw. (IPCAN), Zhejiang University, Hangzhou 310027, China. (e-mail: chongwenhuang@zju.edu.cn ). M. Debbah is with Khalifa University of Science and Technology, P O Box 127788, Abu Dhabi, UAE (email: merouane.debbah@ku.ac.ae). H. Vincent Poor is with the Department of Electrical and Computer Engineering, Princeton University, Princeton, NJ 08544 USA (e-mail: poor@princeton.edu). L. Hanzo is with the School of Electronics and Computer Science, University of Southampton, SO17 1BJ Southampton, U.K. (e-mail: lh@ecs.soton.ac.uk).}}
\maketitle

\begin{abstract}
By integrating a nearly infinite number of reconfigurable elements into a finite space, a spatially continuous array aperture is formed for holographic multiple-input multiple-output (HMIMO) communications. This three-part tutorial aims for providing an overview of the latest advances in HMIMO communications. As Part I of the tutorial, this letter first introduces the fundamental concept of HMIMO and reviews the recent progress in HMIMO channel modeling, followed by a suite of efficient channel estimation approaches. Finally, numerical results are provided for demonstrating the statistical consistency of the new HMIMO channel model advocated with conventional ones and evaluating the performance of the channel estimators. Parts II and III of the tutorial will delve into the performance analysis and holographic beamforming, and detail the interplay of HMIMO with emerging technologies.
\end{abstract}


\section{Introduction}
\IEEEPARstart{T}{he} ever-increasing demand for high throughput and seamless connectivity is propelling the evolution of wireless networks. In order to meet the stringent requirements of emerging applications in the forthcoming sixth-generation (6G) networks, innovative technologies are urgently needed for further enhancing the network capacity and connectivity, while reducing both the latency and energy consumption. In this context, the recent intelligent metasurface technology shows great promise \cite{WC_2020_Huang_Holographic, JWCN_2019_Renzo_Smart, WC_2022_An_Codebook}. In general, metasurfaces are artificial planar structures constructed of a large number of sub-wavelength meta-atoms \cite{Proc_2022_Alexandropoulos_Pervasive}. By appropriately stimulating each meta-atom with the aid of a controller, it is possible to manipulate the electromagnetic (EM) properties of incident waves, such as their amplitude, phase, and polarization, at an unprecedented level \cite{JWCN_2019_Renzo_Smart, WC_2022_An_Codebook}. Over the past few years, energy-efficient reconfigurable intelligent surfaces (RISs) have shown significant potential in creating smart radio environments to enhance the communication quality-of-service (QoS) \cite{TWC_2023_Xu_Antenna, TGCN_2022_An_Joint}.

Beyond serving as passive reflectors, intelligent metasurfaces can also be exploited as reconfigurable antenna arrays, which has led to a recent surge of research in \emph{Holographic Multiple-Input Multiple-Output (HMIMO) Communications\footnote{Note that active metasurfaces at the transceivers are not intended to replace passive RIS, and in fact, these two technologies could be used together in future wireless networks to complement each other.}} \cite{TSP_2018_Hu_Beyond, JSAC_2020_Dardari_Communicating, JSAC_2020_Pizzo_Spatially}. Specifically, by densely packing a nearly infinite number of radiating elements into a finite area, a spatially-continuous EM aperture is formed, which is capable of approaching the ultimate capacity limit of wireless channels \cite{TWC_2022_Pizzo_Fourier}. Through utilizing the radio frequency (RF)-free metasurface technique, HMIMO implements amplitude and phase tuning, as well as signal processing in the EM domain at an appealingly low hardware cost and energy consumption \cite{TSP_2018_Hu_Beyond}. Moreover, as the spacing between radiating elements becomes denser, HMIMO becomes capable of forming a pencil beam having a low sidelobe leakage \cite{WC_2020_Huang_Holographic}. Additionally, an electromagnetically large array leads to an increased Fraunhofer distance, rendering near-field HMIMO communications dominant, which is expected to improve the multiplexing gain in non-scattering environments \cite{JSAC_2020_Dardari_Communicating}.

Despite the above potential benefits, the implementation of HMIMO comes with several new challenges. Specifically, to fully evaluate the theoretical limits and develop effective enabling technologies for HMIMO communications, it is necessary to have tractable channel models that can accurately characterize wave propagation in the EM domain \cite{JSAC_2020_Pizzo_Spatially, TSP_2018_Hu_Beyond}. Compared to their conventional counterparts, the new HMIMO channel models have to consider two crucial factors: \emph{i) the spatially continuous nature of the transceiver aperture}, and \emph{ii) the electromagnetically large size of the transceiver aperture}. This means that HMIMO channels can be mathematically modeled by their EM characteristics, such as the current distributions on the transceiving surfaces. Additionally, the spatial correlation and mutual coupling among densely packed radiating elements cannot be neglected in HMIMO communications \cite{TWC_2022_Pizzo_Fourier}. It is also essential to accurately model the near-field fading, taking into account the propagation distance.

In this Letter, we review the recent advances in HMIMO communications, focusing on channel modeling and estimation. The rest of this Letter is organized as follows. In Section \ref{sec2}, we review both the deterministic and stochastic channel modeling methods of HMIMO systems. Section \ref{sec3} introduces several channel estimation approaches that have been developed. Numerical results are presented in Section \ref{sec4} for validating the accuracy of channel models and for evaluating the performance of channel estimators. Finally, Section \ref{sec5} concludes this Letter.
\section{Channel Modeling of HMIMO Communications}\label{sec2}
\subsection{Deterministic Channel Modeling Method}
Deterministic channels are typically characterized either through ray tracing or actual channel measurements to model the propagation environment \cite{JSAC_2020_Dardari_Communicating}. To characterize the path loss between a pair of closely-spaced large intelligent surfaces (LISs), Hu \emph{et al.} \cite{TSP_2018_Hu_Beyond} proposed a deterministic channel model for extending the classic Friis transmission formula. Furthermore, Williams \emph{et al.} \cite{TWC_2022_Williams_Electromagnetic} proposed a complete narrowband communication model for HMIMO systems, which accounts for three different characteristics, including the propagation and reflection of waves through the waveguides that feed the antenna elements, the mutual coupling through both the air and waveguides, and the insertion losses. More recently, Wang \emph{et al.} \cite{Globecom_2022_Wang_Electromagnetic} proposed an EM-compliant channel model for HMIMO systems. This model can accurately capture both the propagation channel characteristics and the imperfections caused by mutual coupling at the transceivers, such as the antenna pattern distortion.

Note that although deterministic channel models provide accurate predictions of the signal propagation in a given environment, they generally require solving Maxwell’s complex equations and are only applicable to specific scenarios. This makes it challenging to develop a universal framework for performance analysis \cite{TSP_2018_Hu_Beyond, JSAC_2020_Dardari_Communicating}.
\subsection{Stochastic Channel Modeling Method}
Stochastic models are preferred for conducting communication theoretic analysis, since they provide valuable insights into the statistical properties of a class of propagation environments \cite{JSAC_2020_Pizzo_Spatially}. Clarke's classic stochastic model relies on the assumption of an isotropic and scalar wave propagation environment in the far field. In order to overcome this limitation and accurately capture the fundamental properties of wave propagation in wireless communications, Pizzo \emph{et al.} \cite{TWC_2022_Pizzo_Fourier, JSAC_2020_Pizzo_Spatially} developed a small-scale fading model from an EM perspective. This new HMIMO channel model provides a physically meaningful approach to characterizing the spatial correlation that arises from the directionality of arbitrarily distributed scatterers.

\subsubsection{Channel Statistics}
In the far field, the three-dimensional (3D) small-scale fading can be characterized using a spatially stationary correlated Gaussian scalar random field denoted by $h$. For a pair of arbitrary spatial samples having a coordinate difference $\mathbf{\Delta }=\left [ x,y,z \right ]^{T}\in \mathbb{R}^{3}$, the spatial correlation function $r_{h}\left ( \mathbf{\Delta } \right )$ can be expressed as \cite{JSAC_2020_Pizzo_Spatially}
\begin{align}\label{eq4}
 r_{h}\left ( \mathbf{\Delta } \right )=\frac{1}{\left ( 2\pi \right )^{3}}\iiint_{-\infty }^{\infty }S_{h}\left ( \mathbf{k} \right )e^{i\mathbf{k}^{T}\mathbf{\Delta }}dk_{x}dk_{y}dk_{z},
\end{align}
where $S_{h}\left ( \mathbf{k} \right )$ is the power spectral density (PSD) in the wavenumber domain with $\mathbf{k}=\left [ k_{x},k_{y},k_{z} \right ]^{T}\in \mathbb{R}^{3}$.

By imposing a restriction on $h$, namely that of obeying the scalar Helmholtz equation, a physically meaningful PSD function becomes impulsively supported on the surface of a sphere \cite{JSAC_2020_Pizzo_Spatially}, which is expressed as
\begin{align}\label{eq2}
 S_{h}\left ( \mathbf{k} \right )=A_{h}^{2}\left ( \mathbf{k} \right )\delta \left ( k_{x}^{2}+ k_{y}^{2}+k_{z}^{2}-\kappa ^{2} \right ),
\end{align}
where $A_{h}\left ( \mathbf{k} \right )\geq 0$ is the spectral factor that determines the spatial directional properties based on the specific scattering environment, while $\kappa =2\pi /\lambda $ is the wavenumber with $\lambda $ representing the wavelength.

Note that a channel having a PSD defined by \eqref{eq2} is equivalent to Clarke’s model in an isotropic scattering environment, where the multipath components are uniformly strong in all directions and for all antennas. To elaborate, we examine the following two scenarios within this context:
\begin{itemize}
\item \emph{3D Setup:} By considering the normalized power of a 3D isotropic channel $h$, we have $A_{h}\left ( \mathbf{k} \right )=2\pi /\sqrt{\kappa }$ \cite{JSAC_2020_Pizzo_Spatially}. The autocorrelation function (ACF) of $h$ can be derived by performing a 3D inverse Fourier transform of its spatial frequency spectrum, resulting in $r_{h}\left ( \mathbf{\Delta } \right )=\text{sinc}\left ( \kappa \left \| \mathbf{\Delta } \right \|/\pi \right )$. Here, $\textrm{sinc}\left ( x \right )=\sin\left ( \pi x \right )/\left ( \pi x \right )$ is the sinc function and $\left \| \mathbf{\Delta } \right \|$ represents the distance between any pairs of spatial points. It is worth noting that the samples that are spaced apart by an integer multiple of $\lambda /2$ are mutually uncorrelated under the assumption of isotropic small-scale fading \cite{JSAC_2020_Pizzo_Spatially}.
\item \emph{2D Setup:} When examining a linear antenna array at the receiver, the isotropic field $h$ is two-dimensional (2D) and can be calculated over any plane that encompasses the receiving line because of the cylindrical symmetry of the HMIMO channel. By normalizing the channel gain with $A_{h}\left ( \mathbf{k} \right )=2\sqrt{\pi }$, the ACF of $h$ is given by $r_{h}\left ( \mathbf{\Delta } \right )=J_{0}\left ( \kappa \left \| \mathbf{\Delta } \right \| \right )$ \cite{JSAC_2020_Pizzo_Spatially}, where $J_{0}\left ( x \right )$ is the Bessel function of the first kind and order $0$.
\end{itemize}
Therefore, \eqref{eq2} presents a general model for characterizing the small-scale fading over compact HMIMO transceiver arrays and it is also consistent with Clarke's classic models.
\begin{figure}[!t]
\centering
\includegraphics[width=10cm]{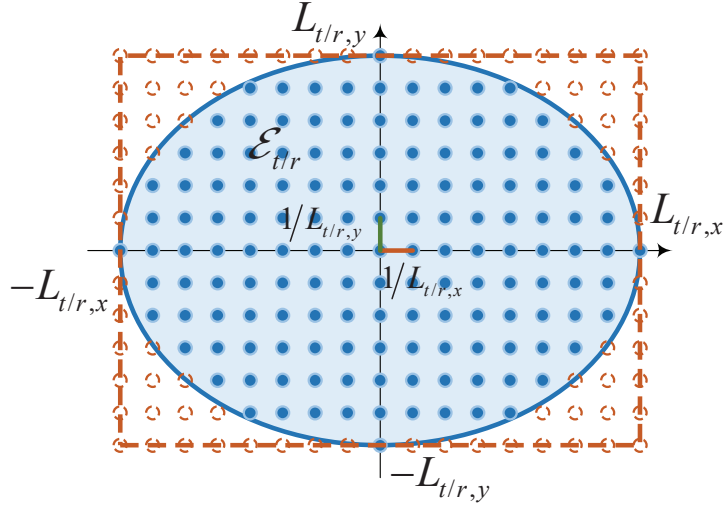}
\caption{The 2D lattice ellipse $\mathcal{E}_{t/r}$ wavenumber spectral support of $\mathbf{H}$.}
\label{fig_1}
\end{figure}
\subsubsection{HMIMO Channel Generation}
Consider a pair of planar arrays oriented vertically and parallel to each other, spanning rectangular regions with dimensions of $\left ( L_{t,x}\lambda \times L_{t,y}\lambda \right )$ and $\left ( L_{r,x}\lambda \times L_{r,y}\lambda \right )$, respectively. The transmit array is equipped with $N_{t}$ antennas, while the receive array has $N_{r}$ antennas. Assuming that the direct path between the transmitter and the receiver is blocked, the HMIMO channel $\mathbf{H}\in \mathbb{C}^{N_{r}\times N_{t}}$ between these two electromagnetically large arrays can thus be modeled using circularly symmetric complex Gaussian (CSCG) correlated random variables, which can be statistically described using the four-dimensional (4D) Fourier plane-wave spectral representation \cite{TWC_2022_Pizzo_Fourier}. By taking a finite number of samples uniformly in the wavenumber domain, the HMIMO channel can be efficiently constructed using the \emph{Fourier plane-wave series expansion}. Specifically, $\mathbf{H}$ can be tightly approximated by
\begin{align}\label{eq3}
 \mathbf{H} =&\sqrt{N_{r}N_{t}}\sum_{\left ( l_{x},l_{y} \right )\in \mathcal{E}_{r}}\sum_{\left ( n_{x},n_{y} \right )\in \mathcal{E}_{t}}H_{a}\left ( l_{x},l_{y},n_{x},n_{y} \right )\notag \\
 &\times \mathbf{a}_{r}\left ( l_{x},l_{y} \right )\mathbf{a}_{t}^{H}\left ( n_{x},n_{y}\right ),
\end{align}
where
\begin{align}
\left [ \mathbf{a}_{t}\left ( n_{x},n_{y} \right ) \right ]_{j}=&\frac{1}{\sqrt{N_{t}}}e^{-j\kappa \left ( t_{x_{j}}n_{x}/L_{t,x}+t_{y_{j}}n_{y}/L_{t,y} \right )}\notag\\
&\times e^{-j\kappa t_{z}\sqrt{ 1-\left ( n_{x}/L_{t,x} \right )^{2}-\left ( n_{y}/L_{t,y} \right )^{2}}},\\
\left [ \mathbf{a}_{r}\left ( l_{x},l_{y} \right ) \right ]_{i}=&\frac{1}{\sqrt{N_{r}}}e^{j\kappa \left ( r_{x_{i}}l_{x}/L_{r,x}+r_{y_{i}}l_{y}/L_{r,y} \right )}\notag\\
&\times e^{j\kappa r_{z}\sqrt{1-\left ( l_{x}/L_{r,x} \right )^{2}-\left ( l_{y}/L_{r,y} \right )^{2}}},
\end{align}
represent the transmit response that connects the impulsive excitation current at a specific point $\mathbf{t}_{j}=\left [ t_{x_{j}},t_{y_{j}},t_{z} \right ]^{T}$ with the propagation direction of the transmitted field, and the receive response that maps the receive direction to the current induced at a particular point $\mathbf{r}_{i}=\left [ r_{x_{i}},r_{y_{i}},r_{z} \right ]^{T}$, respectively. Additionally, the 2D lattice ellipses, i.e., $\mathcal{E}_{t}$ and $\mathcal{E}_{r}$ in \eqref{eq3}, are defined by
\begin{align}
 \mathcal{E}_{t}&=\left \{ \left ( n_{x},n_{y} \right )\in \mathbb{Z}^{2}|\left ( n_{x}/L_{t,x} \right )^{2}+\left ( n_{y}/L_{t,y} \right )^{2}\leq 1 \right \},\\
 \mathcal{E}_{r}&=\left \{ \left ( l_{x},l_{y} \right )\in \mathbb{Z}^{2}|\left ( l_{x}/L_{r,x} \right )^{2}+\left ( l_{y}/L_{r,y} \right )^{2}\leq 1 \right \},
\end{align}
as depicted in Fig. \ref{fig_1}, where the blue and red dots represent the wavenumber harmonics associated with propagating and evanescent waves, respectively. Moreover, the angular response $H_{a}\left ( l_{x},l_{y},n_{x},n_{y} \right )\sim \mathcal{CN}\left ( 0,\sigma ^{2}_{l_{x},l_{y},n_{x},n_{y}} \right )$ characterizes the channel coupling between each transmit-receive direction pairs and it is modeled as a statistically independent CSCG random variable with a mean of zero and a variance of
\begin{align}
 \sigma ^{2}_{l_{x},l_{y},n_{x},n_{y}}=&\frac{1}{\left ( 2\pi \right )^{4}}\iiiint_{\mathcal{W}_{t}\left ( n_{x},n_{y} \right )\times \mathcal{W}_{r}\left ( l_{x},l_{y} \right )}\notag \\
 &\times S\left ( k_{x},k_{y},\kappa _{x},\kappa _{y} \right )dk_{x}dk_{y}d\kappa _{x}d\kappa _{y},
\end{align}
where $S\left ( k_{x},k_{y},\kappa _{x},\kappa _{y} \right )$ represents the band-limited 4D PSD of $\mathbf{H}$, which accounts for the angular selectivity of the scattering. The sets $\mathcal{W}_{t}\left ( n_{x},n_{y} \right )$ and $\mathcal{W}_{r}\left ( l_{x},l_{y} \right )$ represent the $\left ( n_{x},n_{y} \right )$-th transmit direction grid and the $\left ( l_{x},l_{y} \right )$-th receive direction grid, respectively, which are defined in (24) and (25) of \cite{TWC_2022_Pizzo_Fourier}. The size of each angular set is inversely proportional to the antenna array aperture. Note that $\sigma ^{2}_{l_{x},l_{y},n_{x},n_{y}}$ evolve slowly in time compared to the channel coefficients, and can be measured by averaging several channel realizations for optimizing the communication design \cite{TWC_2022_Pizzo_Fourier}.

The HMIMO channel model in \eqref{eq3} captures the fundamental nature of EM propagation in an arbitrary scattering environment and it is applicable in the radiative near-field as well. This is because the wave propagation can always be decomposed into plane waves, regardless of the distance between the transmitter and the receiver as well as the scattering conditions. The accuracy of the channel approximation depends on the number of sampling points, with a larger number of samples leading to a more accurate representation but also resulting in increased complexity \cite{JSAC_2020_Pizzo_Spatially}.

Recently, Wei \emph{et al.} \cite{JSTSP_2022_Wei_Multi} extended the Fourier plane-wave series expansion model for accommodating multi-user HMIMO scenarios. Additionally, Demir \emph{et al.} \cite{WCL_2022_Demir_Channel} utilized conventional antenna array responses and angular spread functions for characterizing multi-path components and phase distributions, and developed an HMIMO channel model relying on a closed-form expression for the spatial correlation matrix.
\section{Channel Estimation for HMIMO Systems}\label{sec3}
Having accurate channel state information (CSI) is crucial for recovering signals at the receiver. Unfortunately, channel estimation becomes extremely challenging in HMIMO systems due to the use of electromagnetically large surfaces \cite{WCL_2022_Demir_Channel, TCOM_2022_An_Low}. Next, we will introduce several efficient channel estimation techniques tailored for single-user scenarios, which can be readily adapted for multi-user scenarios by assigning orthogonal pilot sequences to each user. For practical multiuser HMIMO communications, it may be necessary to determine the appropriate pilot reuse factor and design the pilot pattern to strike a favorable tradeoff between the overhead and the achievable performance.

Specifically, we consider the uplink of an HMIMO system, where the base station (BS) is equipped with a uniform planar array (UPA) consisting of $N$ antennas. Let $\mathbf{h}\in \mathbb{C}^{N\times 1}$ represent the channel from the single-antenna user to the BS, which is modeled by correlated Rayleigh fading having a spatial correlation matrix of $\mathbf{R}\in \mathbb{C}^{N\times N}$ satisfying $\textrm{tr}\left ( \mathbf{R} \right )=N\beta $, where $\beta$ represents the average channel gain accounting for both pathloss and shadowing effects. During the uplink training phase, the user transmits a predefined pilot sequence, while the signal received by the BS can be expressed as
\begin{align}
 \mathbf{y}=\sqrt{\gamma }\mathbf{h}+\mathbf{n},
\end{align}
where $\gamma > 0$ is the pilot signal-to-noise-ratio (SNR) and $\mathbf{n}\sim \mathcal{CN}\left ( \mathbf{0},\mathbf{I}_{N} \right )$ is the additive white Gaussian noise. If there is no prior information regarding the channel statistics or array geometry, the least-squares (LS) estimate of $\mathbf{h}$ is given by
\begin{align}
 \widehat{\mathbf{h}}_{\text{LS}}=\mathbf{y}/\sqrt{\gamma }.
\end{align}
Alternatively, if the spatial correlation matrix $\mathbf{R}$ is known, we can obtain the minimum mean-squared error (MMSE) estimate of $\mathbf{h}$ using:
\begin{align}\label{eq9}
 \widehat{\mathbf{h}}_{\text{MMSE}}=\sqrt{\gamma }\mathbf{R}\left ( \gamma \mathbf{R}+\mathbf{I}_{N} \right )^{-1}\mathbf{y}.
\end{align}

\subsection{Subspace-Based Channel Estimation}
The dense deployment of antennas in a finite space would result in highly rank-deficient spatial correlation matrices. In particular, the rank of the spatial correlation matrix $\mathbf{R}_{\text{iso}}$ of an HMIMO channel under isotropic scattering is approximately equal to $\text{rank}\left ( \mathbf{R}_{\text{iso}} \right ) \approx \pi N \Delta^{2} /\lambda^{2} $, where $\Delta$ represents the element spacing \cite{TWC_2022_Pizzo_Fourier}. When $\Delta =\lambda /4$, almost $80\%$ of the eigenvalues are zero. The grade of rank deficiency increases as the antenna spacing decreases and it becomes more pronounced for non-isotropic scattering. Taking advantage of the rank deficiency caused by the array geometry, Demir \emph{et al.} \cite{WCL_2022_Demir_Channel} proposed a novel subspace-based channel estimation approach. This technique identifies a subspace of reduced rank that covers the eigenspace of any spatial correlation matrix and outperforms the conventional LS estimator without requiring any information about the channel statistics.

Specifically, let $\text{rank}\left ( \mathbf{R} \right ) = r$ denote the rank of $\mathbf{R}$, where $1\leq r\leq N$. The compact eigenvalue decomposition of $\mathbf{R}$ is denoted by $\mathbf{R}=\mathbf{U}\mathbf{\Lambda }\mathbf{U}^{H}$, where $\mathbf{\Lambda }\in \mathbb{C}^{r\times r}$ is a diagonal matrix that contains the non-zero eigenvalues, and the columns of $\mathbf{U}\in \mathbb{C}^{N\times r}$ are the corresponding orthonormal eigenvectors. Hence, the channel $\mathbf{h}$ is represented as $\mathbf{h}=\mathbf{R}^{1/2}\mathbf{e}=\mathbf{U}\mathbf{\Lambda} ^{1/2}\mathbf{e}$, where $\mathbf{e}\sim \mathcal{CN}\left ( \mathbf{0},\mathbf{I}_{r} \right )$. This means that all possible channel realizations exist in the subspace spanned by $\mathbf{U}$ \cite{WCL_2022_Demir_Channel}. If only the subspace spanned by $\mathbf{U}$ is known, one can use the reduced-subspace LS (RS-LS) estimator, which is defined as:
\begin{align}\label{eq12}
 \widehat{\mathbf{h}}_{\text{RS-LS}}=\mathbf{U}\mathbf{U}^{H}\mathbf{y}/\sqrt{\gamma }.
\end{align}
Furthermore, $\widehat{\mathbf{h}}_{\text{RS-LS}}$ approaches $\widehat{\mathbf{h}}_{\text{MMSE}}$ as $\gamma \rightarrow \infty$, which implies that the performance of the RS-LS method is comparable to that of the MMSE method in high-SNR scenarios. Compared to the plain LS method, the RS-LS method removes noise from $\left ( N-r \right )$ dimensions by projecting the received signal onto a reduced subspace. As a result, the effective SNR is increased by a factor of $N/r$.

To avoid the need for knowing $\mathbf{R}$, one can implement the RS-LS method using a different correlation matrix $\overline{\mathbf{R}}$ that captures the general array geometry. For instance, by setting $\overline{\mathbf{R}} = \mathbf{R}_{\text{iso}}$, the subspace spanned by $\overline{\mathbf{R}}$ will contain the subspace spanned by any other $\mathbf{R}$ for a given array geometry \cite{WCL_2022_Demir_Channel}. For any $\overline{\mathbf{R}}$ ensuring that the subspace spanned by it contains the subspace spanned by $\mathbf{R}$, the conservative RS-LS estimator is given by
\begin{align}
 \widehat{\mathbf{h}}_{\text{RS-LS, con}}=\overline{\mathbf{U}}\overline{\mathbf{U}}^{H}\mathbf{y}/\sqrt{\gamma },
\end{align}
where the columns $\overline{\mathbf{U}}\in \mathbb{C}^{N\times \overline{r}}$ are the orthonormal eigenvectors that correspond to the $\overline{r}$ non-zero eigenvalues of $\overline{\mathbf{R}}$.

\subsection{Sparse Channel Estimation}
Conventional methods of estimating HMIMO channels often result in an excessive pilot overhead because of the large number of closely spaced elements at both ends \cite{WCL_2022_Demir_Channel}. Fortunately, the EM channels of most practical communication scenarios can be sparsely represented in the angular domain, making it possible to use compressed sensing (CS) techniques for reducing the pilot overhead. In \cite{TCOM_2021_Wan_Terahertz}, Wan \emph{et al.} developed a two-stage channel estimation approach for estimating the broadband TeraHertz (THz)-HMIMO channels with the aid of RISs. In the downlink transmission, the RIS applies spatial bandpass filtering to help users obtain a rough estimate of the line-of-sight (LoS) angles. Then, the users having similar LoS angles are grouped together for uplink transmission, and the coarsely estimated LoS angles are exploited for enhancing the channel estimation performance. Furthermore, to reduce the uplink pilot overhead, a CS-based channel estimator was developed by exploiting the dual sparsity of THz-HMIMO channels in both the angular and delay domains.

Due to the spherical wavefront of near-field HMIMO communications, the channel sparsity in the angular domain may no longer be applicable. To address this issue, Cui and Dai \cite{TCOM_2022_Cui_Channel} exploited the polar-domain sparsity instead. Explicitly, they proposed an on-grid polar-domain simultaneous orthogonal matching pursuit (OMP) algorithm that considers both the angular and distance information for efficiently estimating the near-field channel. Additionally, a parametric channel estimation approach was proposed in \cite{arXiv_2021_Ghermezcheshmeh_Parametric} for LoS-dominated HMIMO communications. This method models the far-field channel using path parameters and exploits the radiated pattern structure for estimating the channel parameters at a substantially reduced training overhead. In the near field, the channel was modeled as a superposition of channels through individual tiles created by partitioning the continuous aperture. These individual channels can be estimated using a similar approach as in the far field.

In a nutshell, it is crucial to develop efficient channel estimation techniques based on new channel models to fully unlock the potential of HMIMO communications. The HMIMO channel heavily depends on the specific geometric configuration, such as the relative positioning and orientation of transceiver surfaces. This presents a unique opportunity for simultaneously tackling both channel estimation and direct localization.

\section{Numerical Results}\label{sec4}
This section presents our numerical results for verifying the accuracy of the channel models and of the channel estimators.
\subsection{Accuracy of the Plane-Wave Series Expansion Model}
We first validate the accuracy of the Fourier plane-wave series expansion model. For simplicity, we only consider the scenario of isotropic propagation. In this case, the HMIMO channel is generated by collecting $N$-dimensional uniform spatial samples into a random vector $\mathbf{h}\in \mathbb{C}^{N\times 1}$. We assess the accuracy of the Fourier plane-wave series expansion method by comparing it with the conventional spatially-stationary random field channels $\mathbf{h} = \mathbf{R}^{1/2}\mathbf{e}$, where $\mathbf{e}\sim \mathcal{CN}\left ( \mathbf{0},\mathbf{I}_{N} \right )$, and the spatial correlation matrix $\mathbf{R}$ is obtained by sampling Clarke’s ACF. By choosing a uniform spacing $\Delta $ along both the $x$- and $y$-axes, $\mathbf{R}$ becomes a symmetric Toeplitz matrix that is fully characterized by its first row. Hence, we only plot the first row of the spatial correlation matrices.

\begin{figure}[!t]
\centering
\subfloat[]{\includegraphics[width=8cm]{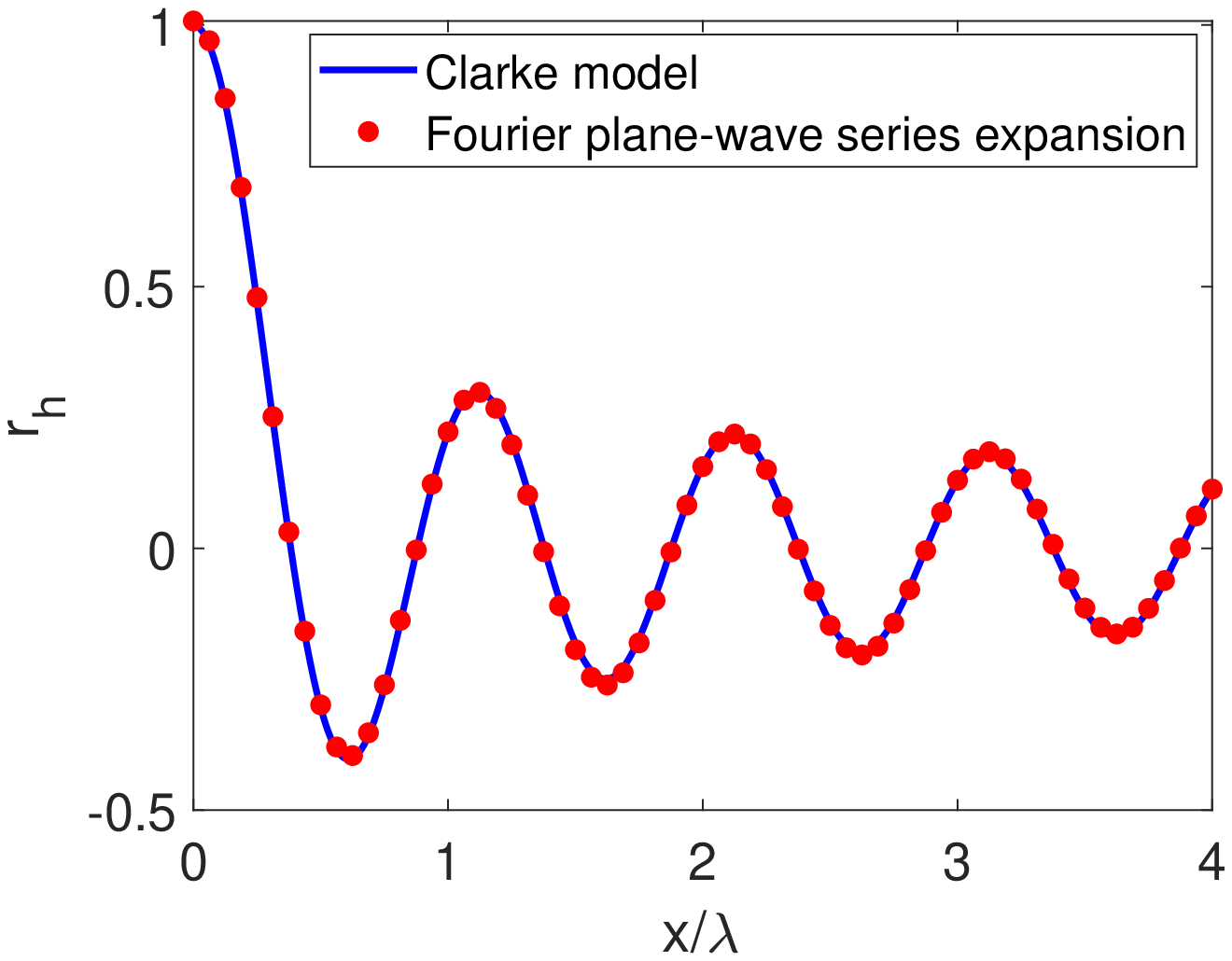}%
\label{fig_2a}}
\subfloat[]{\includegraphics[width=8cm]{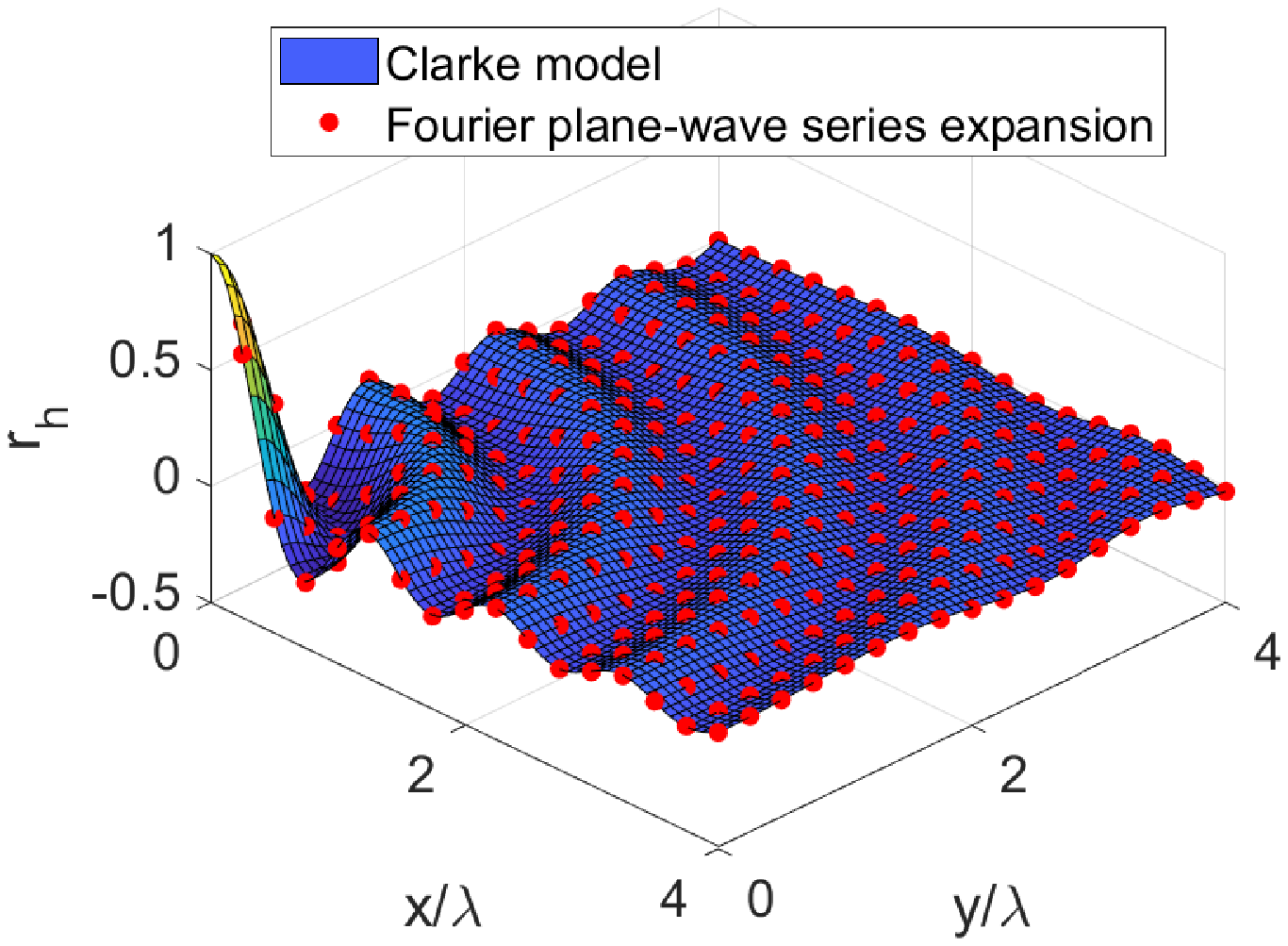}%
\label{fig_2b}}
\caption{(a) The 1D ACF of $h$ as a function of $x/\lambda \in \left [ 0,L_{x}/4 \right ]$ with $L_{x}=16 $ and $\Delta _{x}=\lambda /16$. (b) The 2D ACF of $h$ as a function of $x/\lambda \in \left [ 0,L_{x}/4 \right ]$ and $y/\lambda \in \left [ 0,L_{y}/4 \right ]$ with $L_{x}= L_{y} =16 $ and $\Delta _{x} = \Delta _{y} =\lambda /4$.}
\label{fig_2}
\end{figure}

Fig. \ref{fig_2a} illustrates the one-dimensional (1D) ACF of numerically generated samples $h$ employing a linear aperture of $L_{x}=16$ and sampling spacing of $\Delta_{x} =\lambda /16$. Note that the empirical ACF closely matches its closed form, i.e., $r_{h}\left ( x_{n} \right )=J_{0}\left ( \kappa x_{n} \right )$. Furthermore, Fig. \ref{fig_2b} plots the 2D ACF of $h$ over a rectangular grid having side lengths of $L_{x}=L_{y}=16 $ with uniform spacing $\Delta_{x} =\Delta_{y} =\lambda /4$. We can observe that the Fourier plane-wave series expansion method accurately models the HMIMO channel of compact rectangular arrays having practical sizes.
\begin{figure}[!t]
\centering
\subfloat[]{\includegraphics[width=8cm]{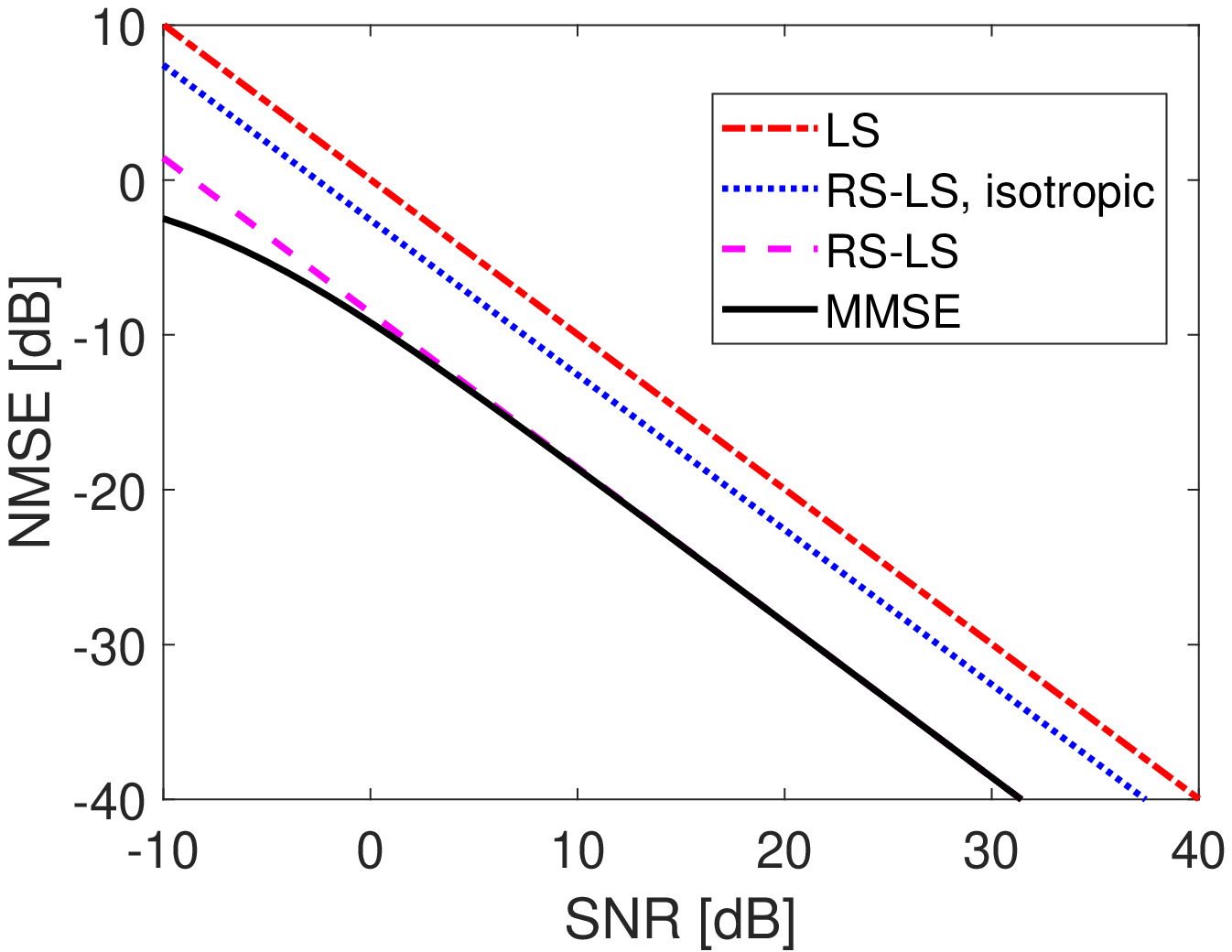}%
\label{fig_3a}}
\subfloat[]{\includegraphics[width=8cm]{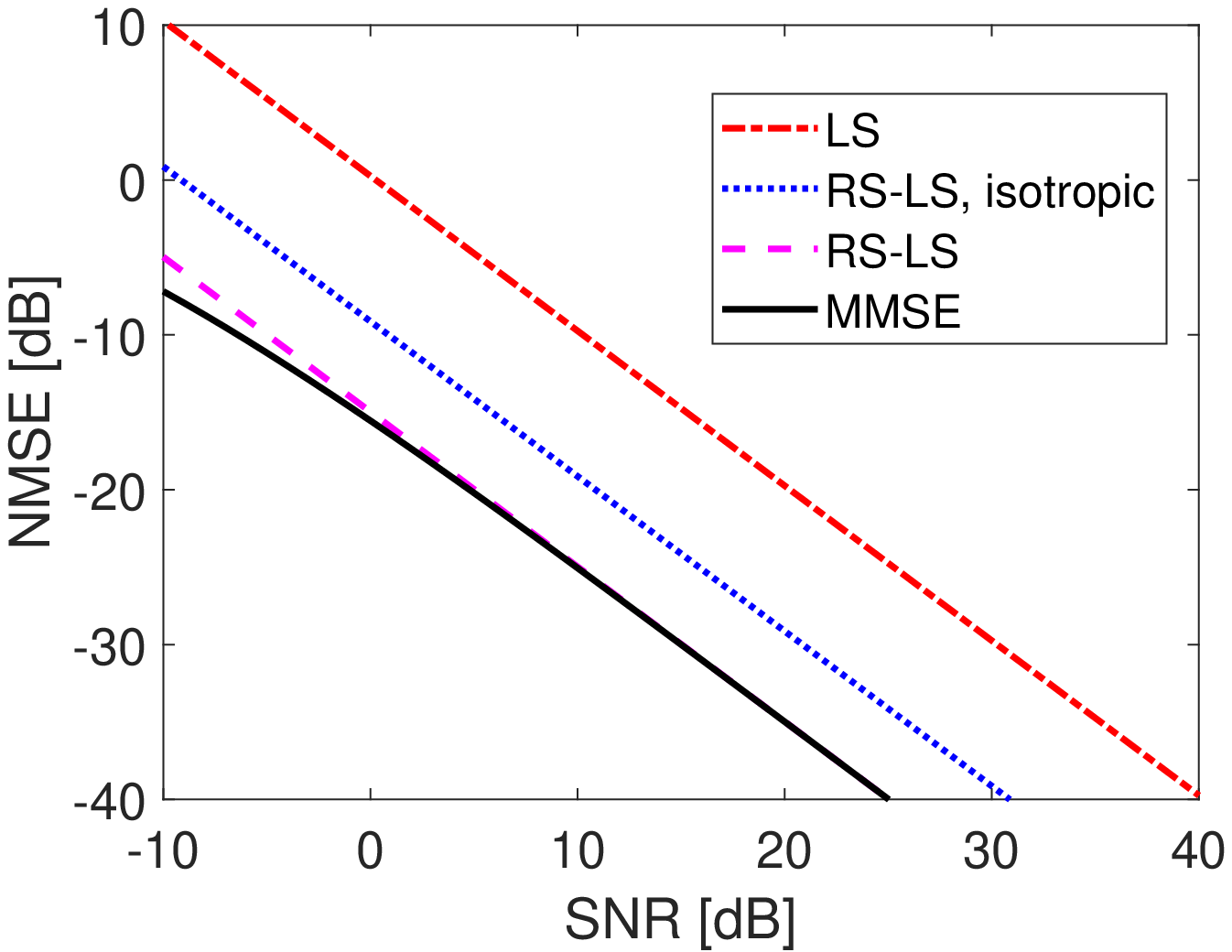}%
\label{fig_3b}}
\caption{NMSE versus SNR for the $32\times 32$ UPA with (a) $\Delta =\lambda /4$; (b) $\Delta =\lambda /16$.}
\label{fig_3}
\end{figure}
\subsection{Performance Evaluation of Channel Estimators}
Next, we will evaluate the performance of the channel estimation methods using the normalized mean-square error (NMSE) as a metric. For simplicity, we consider a non-isotropic scenario by truncating the first $\textrm{rank}\left ( \mathbf{R}_{\text{iso}} \right )/4$ eigenvalues of $\mathbf{R}_{\text{iso}}$. The average channel gain is set to $\beta = 1$, and a $\left (32\times 32\right )$ UPA is employed. Fig. \ref{fig_3a} shows the NMSE versus SNR for an antenna spacing of $\Delta =\lambda /4$. Note that the MMSE estimator provides the lowest NMSE, while the statistics-agnostic LS estimator has an $8.5$ dB higher NMSE. The RS-LS estimator utilizes the subspace spanned by the true spatial correlation matrix for eliminating the noise from the null space of the correlation matrix. At low SNR values, RS-LS is inferior to MMSE. However, the gap between RS-LS and MMSE diminishes, as the SNR increases. Note that the ``RS-LS, isotropic'' estimator, which utilizes the array geometry and isotropic subspace, provides a performance gain of $2.5$ dB over LS. The gap between RS-LS and ``RS-LS, isotropic'' is because the true correlation matrix has a lower rank than $\mathbf{R}_{\textrm{iso}}$.

In Fig. \ref{fig_3b}, we reduce the antenna spacing to $\Delta =\lambda /16$. As the array becomes denser, the ratio of the effective rank of the spatial correlation matrix to $N$ is reduced. As a consequence, the NMSE gap between ``RS-LS, isotropic'' and LS estimators has increased to $9$ dB. This gap widens as the rank deficiency induced by the array geometry becomes more significant. It is important to note that the ratio of the rank of the spatial correlation matrix to $N$ is $\pi \Delta ^{2}/\lambda ^{2}$ when $N$ is large and $\Delta $ is small, which increases with $\Delta $, regardless of $N$. Hence, when more antennas are incorporated into a given array size, the performance of the RS-LS estimator improves.

\section{Conclusions}\label{sec5}
In this Letter, we have presented a comprehensive overview of the basic principles of HMIMO communications for next-generation networks. We have reviewed the latest breakthroughs in HMIMO channel modeling and effective channel estimation techniques. The spatially continuous transceiver aperture enables the modeling of HMIMO channels in the EM domain. Finally, the efficacy of the channel models and estimators developed has been validated through numerical simulations. Part II of this Letter will delve deeper into the performance analysis of HMIMO communications, while Part III will identify some open challenges and investigate the interplay between HMIMO and a suite of emerging technologies.
\bibliographystyle{IEEEtran}
\bibliography{ref}
\end{document}